\documentclass[aps,prl,twocolumn,groupedaddress,showpacs,superscriptaddress]{revtex4-1}
\usepackage{graphicx}
\usepackage{amsmath}
\usepackage{color}
\usepackage{float}
\usepackage{epstopdf}
\usepackage{soul,xcolor}
\usepackage[utf8]{inputenc}
\usepackage[colorlinks=true,citecolor=blue,linkcolor=magenta]{hyperref}
\usepackage{verbatim}
\usepackage{color}

\begin{document}
	
	\title{Experimental Realization of Two Qutrits Gate with Tunable Coupling in Superconducting Circuits}
	
	\author{Kai Luo}
	\affiliation{Department of Physics, Harbin Institute of Technology, Harbin 150001, China}
	\affiliation{Department of Physics, Southern University of Science and Technology, Shenzhen 518055, China}
	\affiliation{Shenzhen Institute for Quantum Science and Engineering, Southern University of Science and Technology, Shenzhen 518055, China}
	\author{Wenhui Huang}
	\author{Ziyu Tao}
	\author{Libo Zhang}
	\author{Yuxuan Zhou}
	\affiliation{Department of Physics, Southern University of Science and Technology, Shenzhen 518055, China}
	\affiliation{Shenzhen Institute for Quantum Science and Engineering, Southern University of Science and Technology, Shenzhen 518055, China}
	
	\author{Ji Chu}
	\affiliation{Shenzhen Institute for Quantum Science and Engineering, Southern University of Science and Technology, Shenzhen 518055, China}

	\author{Wuxin Liu}
	\affiliation{Huawei 2012 lab}
	\author{Biying Wang}
	\affiliation{Huawei 2012 lab}
	\author{Jiangyu Cui}
	\affiliation{Huawei 2012 lab}
	
	\author{Song Liu}
	\author{Fei Yan}
	\affiliation{Shenzhen Institute for Quantum Science and Engineering, Southern University of Science and Technology, Shenzhen 518055, China}
	\affiliation{Guangdong Provincial Key Laboratory of Quantum Science and Engineering, Southern University of Science and Technology, Shenzhen, 518055, China}
	\affiliation{Shenzhen Key Laboratory of Quantum Science and Engineering, Southern University of Science and Technology, Shenzhen 518055, China}
	\affiliation{International Quantum Academy, Shenzhen, Guangdong, China}

	\author{Man-Hong Yung}
    \email{yung@sustech.edu.cn }
    \affiliation{Department of Physics, Southern University of Science and Technology, Shenzhen 518055, China}
	\affiliation{Shenzhen Institute for Quantum Science and Engineering, Southern University of Science and Technology, Shenzhen 518055, China}
	\affiliation{Huawei 2012 lab}
	\affiliation{Guangdong Provincial Key Laboratory of Quantum Science and Engineering, Southern University of Science and Technology, Shenzhen, 518055, China}
	\affiliation{Shenzhen Key Laboratory of Quantum Science and Engineering, Southern University of Science and Technology, Shenzhen 518055, China}
	\affiliation{International Quantum Academy, Shenzhen, Guangdong, China}

	\author{Yuanzhen Chen}
	\email{chenyz@sustech.edu.cn}
	\affiliation{Department of Physics, Southern University of Science and Technology, Shenzhen 518055, China}
	\affiliation{Shenzhen Institute for Quantum Science and Engineering, Southern University of Science and Technology, Shenzhen 518055, China}
	\affiliation{Guangdong Provincial Key Laboratory of Quantum Science and Engineering, Southern University of Science and Technology, Shenzhen, 518055, China}
	\affiliation{Shenzhen Key Laboratory of Quantum Science and Engineering, Southern University of Science and Technology, Shenzhen 518055, China}
	\affiliation{International Quantum Academy, Shenzhen, Guangdong, China}
	
	\author{Tongxing Yan}
	\email{yantx@sustech.edu.cn}
	\affiliation{Shenzhen Institute for Quantum Science and Engineering, Southern University of Science and Technology, Shenzhen 518055, China}
	\affiliation{Guangdong Provincial Key Laboratory of Quantum Science and Engineering, Southern University of Science and Technology, Shenzhen, 518055, China}
	\affiliation{Shenzhen Key Laboratory of Quantum Science and Engineering, Southern University of Science and Technology, Shenzhen 518055, China}
	\affiliation{International Quantum Academy, Shenzhen, Guangdong, China}
	
	\author{Dapeng Yu}
	\affiliation{Department of Physics, Southern University of Science and Technology, Shenzhen 518055, China}
	\affiliation{Shenzhen Institute for Quantum Science and Engineering, Southern University of Science and Technology, Shenzhen 518055, China}
	\affiliation{Guangdong Provincial Key Laboratory of Quantum Science and Engineering, Southern University of Science and Technology, Shenzhen, 518055, China}
	\affiliation{Shenzhen Key Laboratory of Quantum Science and Engineering, Southern University of Science and Technology, Shenzhen 518055, China}
	\affiliation{International Quantum Academy, Shenzhen, Guangdong, China}

	\date{\today }

    \begin{abstract}
    Gate-based quantum computation has been extensively investigated using quantum circuits based on qubits. In many cases, such qubits are actually made out of multilevel systems but with only two states being used for computational purpose. While such a strategy has the advantage of being in line with the common binary logic, it in some sense wastes the ready-for-use resources in the large Hilbert space of these intrinsic multi-dimensional systems. Quantum computation beyond qubits (e.g., using qutrits or qudits) has thus been discussed and argued to be more efficient than its qubit counterpart in certain scenarios. However, one of the essential elements for qutrit-based quantum computation, two-qutrit quantum gate, remains a major challenge. In this work, we propose and demonstrate a highly efficient and scalable two-qutrit quantum gate in superconducting quantum circuits. Using a tunable coupler to control the cross-Kerr coupling between two qutrits, our scheme realizes a two-qutrit conditional phase gate with fidelity $89.3\%$ by combining simple pulses applied to the coupler with single-qutrit operations. We further use such a two-qutrit gate to prepare an EPR state of two qutrits with a fidelity of $95.5\%$. Our scheme takes advantage of a tunable qutrit-qutrit coupling with a large on:off ratio. It therefore offers both high efficiency and low cross talk between qutrits, thus being friendly for scaling up. Our work constitutes an important step towards scalable qutrit-based quantum computation. 
    \end{abstract}
    
	\vskip 0.5cm
	\maketitle
	
	
    The past two decades have witnessed an extraordinary progress in our capability of precisely controlling various quantum systems for quantum information processing (QIP). As a milestone, for example, the concept of quantum advantage has been demonstrated recently in different scenarios \cite{arute2019quantum,Wu2021prl,ZHU2021,Zhong21PRL}. Most existing research on QIP, both theoretical and experimental, relies heavily on two-level quantum systems, i.e., qubits, analogue of classical binary bits \cite{nielsen_chuang_2010}. However, it has been suggested that QIP using multi-level systems, such as qutrits (three levels) or qudits ( more than three levels), may have significant advantages over qubit-based schemes due to the added access to extra dimensions of the Hilbert space \cite{PhysRevAKlimov,lanyon2009simplifying,luo2014universal,gokhale2019asymptotic,Kiktenko20pra,wang2020qudits}. Among others are more efficient coding for quantum error correction \cite{Campbell14prl}, more flexible simulation of quantum dynamics \cite{PhysRevXBlok,Neeley_science}, magic state distillation with higher fidelity \cite{Campbellprx14}, and more robust quantum cryptography protocols \cite{,Bechmann00prl,bru02prl,hu2018beating}. These advantages would be particularly welcome in the noisy intermediate-scale quantum era where hardware efficiency is of critical importance \cite{Preskill2018}. Some initial efforts of using qutrits or qudits for QIP include construction of quantum gates and generation of entanglement in photonic systems \cite{krenn2014generation,Babazadeh2017,erhard2018experimental,imany2019high,erhard2020advances,Chi2022} and superconducting quantum circuits \cite{PhysRevXBlok}. These works have already shown encouraging potential applications of multilevel systems in QIP. Further progress requires viable schemes that have good scalability with relatively simple engineering. 

    Circuit quantum electrodynamics (cQED) is the study of the interaction between superconducting circuits and quantized electromagnetic fields in the microwave frequency region \cite{Blais04pra,RevModPhysBlais21}. Due to the diversity of superconducting quantum devices and flexibility in circuit design, as well as a constantly improving understanding of decoherence in these systems \cite{krantz2019a,somoroff2021millisecond}, superconducting quantum circuits based on cQED (especially the mainstream architecture with transmon qubits \cite{Koch07pra}) have nowadays become a leading experimental platform for quantum computation \cite{arute2019quantum,chen2021exponential,krinner2021realizing}. While superconducting qubits, including transmons, are intrinsically multilevel systems, 
    most work only utilize two of them to encode quantum information, and the rest are only exploited as ancillary during the logic gate operation or treated as leakage channel \cite{Motzoi09prl,fedorov2012implementation,barends2014superconducting,Rosenblum18sc,Tongxing19prl}.

    Recently, significant progress has been made towards using transmon as qutrits for QIP, including a proof-of-principle of entanglement generation \cite{PhysRevXBlok,cerveralierta2021} and quantum gates \cite{Yurtalan20prl,Morvan2021PRL,Kononenko2021prr}. In these works, two-qutrit gate operations involving two transmons are realized based on a small residual cross-Kerr coupling between them, results in slow gate operation and correlation error during idle period. Therefore, scheme for two-qutrit gates featuring a stronger coupling with a large on-off ratio are highly desirable to further improve qutrit-based quantum operation fidelity meanwhile maintain scalability. 

    In this work, we develop a scalable two-qutrit conditional phase (Cphase) gate using two transmons of fixed frequencies connected to a tunable coupler. The coupler itself is also a transmon but with a tunable energy spacing, which can be used to vary the effective cross-Kerr coupling between the other two transmons connected to it. Such a coupling scheme has been used to construct conditional phase gates of qubits and is highly favored for its  scalability \cite{Yan2018,Yuan2020prl,Collodo2020prl,Foxen2020prl,Sung2021prx,Stehlik2021prl,Xu_2021}. In our study, the cross-Kerr coupling among higher levels of the two qutrits can be tuned in a broad range from about 100 kHz (idling regime for single-qutrit operations) to around 100 MHz (accumulating conditional phase for the two-qutrit gate) by adjusting the frequency of the coupler, achieving a large on/off ratio. Using such a setup, we demonstrate a two-qutrit Cphase gate with a fidelity of $89.3(\pm1.9)\%$ determined by a quantum process tomography. We note that compared to the qubit case, qutrit quantum gates utilize more and higher energy levels and thus require for more complex control schemes and are also more susceptible to decoherence \cite{PhysRevXBlok}. With a better understanding and engineering of the transmon-coupler setup \cite{chuji2021,Place2021,wang2022towards}, qutrit quantum gates with much improved fidelity can be realized.   

    \begin{figure}
    \includegraphics[width=0.45\textwidth]{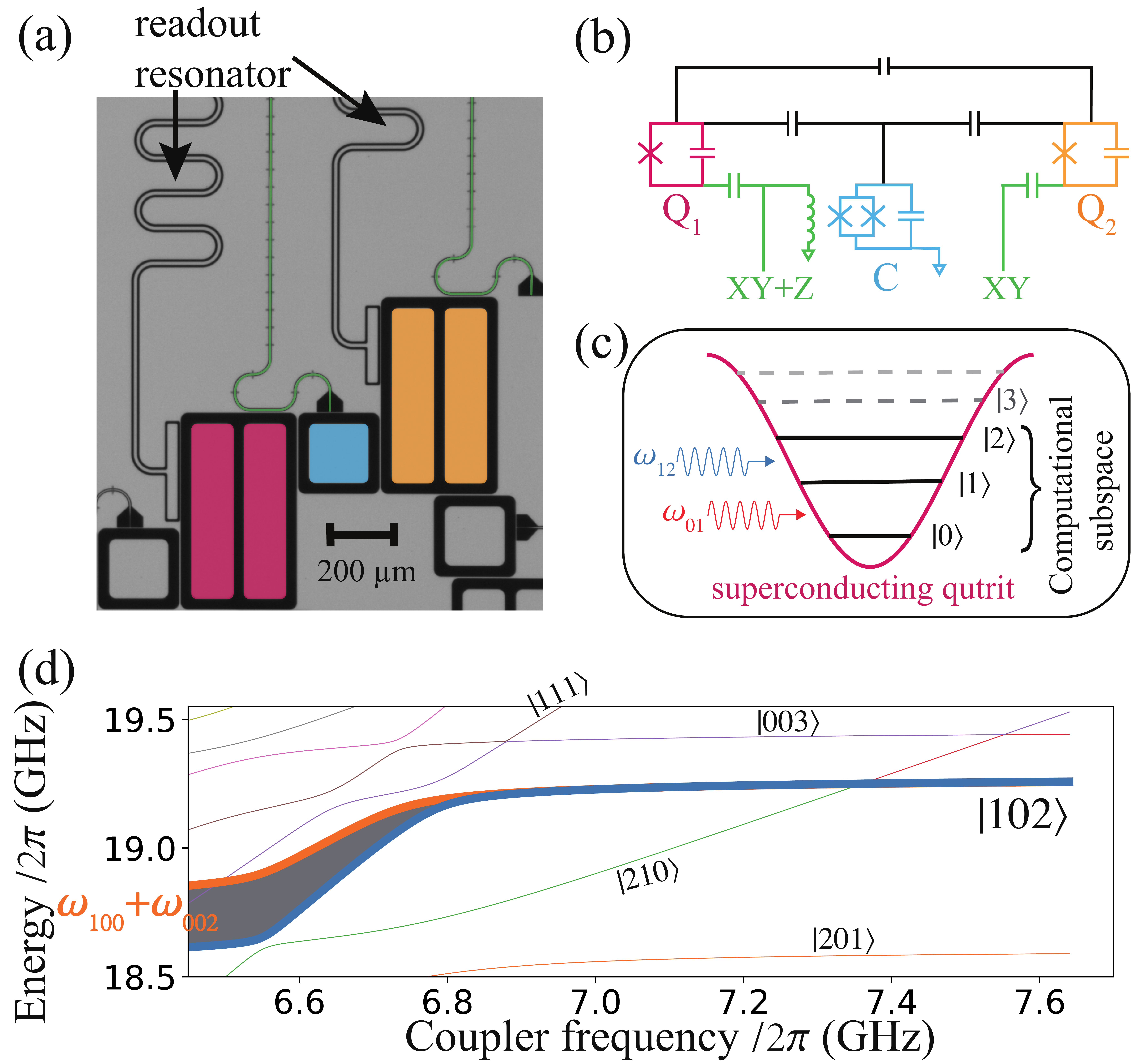}
    \caption{\label{fig:devices}Schematics of device and two-qutrit Cphase gate. (a) False-colored micrograph of the superconducting circuit used in this work. Two transmon qutrits (pink and orange) with fixed frequencies are connected to another transmon with tunable frequency serving as a coupler (blue). Green lines are for control. The left control line is shared by the left transmon for its single-qutrit operations and the coupler for tuning its frequency. Each qutrit is capacitively coupled to a dedicated $\lambda/4$ resonator for readout. (b) Simplified circuit model of the device in (a). (c) Energy diagram of a typical transmon as a multi-level system. The lowest three levels ($|0\rangle,|1\rangle,|2\rangle$) span the computational subspace of a single qutrit. (d) Partial energy diagram of the qutrit-coupler-qutrit system vs coupler frequency. The eigenenergy of state $|102\rangle$ (dark blue curve), $\Tilde{\omega}_{102}$, is compared to the summation (orange) of eigenenergies of the states $|100\rangle$ and $|002\rangle$. Their difference (shaded area) represents the conditional phase $\chi_{102}$, which increases significantly when the coupler is tuned from its sweet point (about $7.64$ GHz) to the frequency range of the qutrits (about $6.6$ GHz). Similar behavior (not shown) is observed for $\chi_{101},\chi_{201}$, and $\chi_{202}$.}
    \end{figure}

     Our superconducting quantum processor consists of eight fixed-frequency transmons, arranged in a ring geometry with nearest neighboring transmons interconnected via frequency-tunable couplers. The two transmons and their coupler used in this study is shown in Fig.~\ref{fig:devices}(a) and its simplified quantum circuit is sketched in Fig.~\ref{fig:devices}(b). Each transmon has a dedicated readout resonator for multiplexed dispersive readout. The tunable coupler consists of a superconducting pad capacitively coupled to the ground via a superconducting quantum interference device (SQUID). Diplexed signals for controlling both the qutrit ($4–8$ $\text{GHz}$) and the coupler (DC-1 $\text{GHz}$) are synthesized at room temperature and transmitted to the device via a shared control line. Such a design substantially reduces wiring complexity and is friendly for scalability. More details about the whole device and the experimental setup can be found in Ref.\cite{chu2021scalable}.

    Typical parameter settings of a transmon guarantee that it has a good coherence performance while maintaining a sufficiently large anharmonicity, which enables it to be operated as not only a qubit but also a higher-dimensional qutrit or qudit \cite{Bianchetti2010prl,Peterer2015prl}. In our study, the lowest three energy levels (labeled as $\vert 0\rangle$, $\vert 1\rangle$, and $\vert 2\rangle$) of a transmon are encoded as logical states of a qutrit (Fig.~\ref{fig:devices}(c)). An arbitrary single-qutrit gate operation can be decomposed into operations in the two subspaces of ($\vert 0\rangle,\vert 1\rangle$) and ($\vert 1\rangle,\vert 2\rangle$) \cite{PhysRevXBlok,Morvan2021PRL}. However, in order to realize single-qutrit gates of high fidelity, the AC Stark effect due to a small anharmonicity of transmon should be taken into account \cite{Liu2016pra,PhysRevXBlok}. For example, a microwave drive resonant to the subspace of ($\vert 0\rangle,\vert 1\rangle$) induces an effective Z rotation in the subspace of ($\vert 1\rangle,\vert 2\rangle$), and vice versa. 

    The system shown in Fig.~\ref{fig:devices} can be effectively described by a model of three coupled nonlinear oscillators. When only the low lying energy levels are considered, its Hamiltonian can be written as \cite{Yan2018} ($\hbar \equiv 1$):
        \begin{eqnarray}
        H=&\sum_{i=1,2,c}(\omega_i a_i^\dagger a_i +\frac{\alpha_i}{2} a_i^\dagger a_i^\dagger a_i a_i)\nonumber\\
        &+\sum_{i\ne j}g_{ij}(a_i^\dagger + a_i)(a_j^\dagger + a_j)
        \label{eq1}
        \end{eqnarray}
    where subscripts 1,2 and C refer to qutrit 1 ($Q_1$), qutrit 2 ($Q_2$), and the tunable coupler ($C$), respectively. $\omega_1/2\pi=6.074$ GHz ($\omega_2/2\pi=6.725$ GHz) is the transition frequency between the ground and the first exited states of $Q_1$($Q_2$), and $\alpha_1/2\pi=-0.256$ GHz ($\alpha_2/2\pi=-0.236$ GHz) is the corresponding anharmonicity. The maximum frequency and anharmonicity of the coupler with a symmetric SQUID are 7.640 and -0.310 GHz, respectively. The three coupling constants determined by the mutual capacitances are designed to be roughly $g_{1C}/2\pi\sim g_{2C}/2\pi$ ($\approx$ 100 MHz) $\gg g_{12}/2\pi$ ($\approx$ 5 MHz), which results in a better adiabatic control and a small residual effective cross-Kerr interaction.
    
    \begin{figure}
    \includegraphics[width=0.47\textwidth]{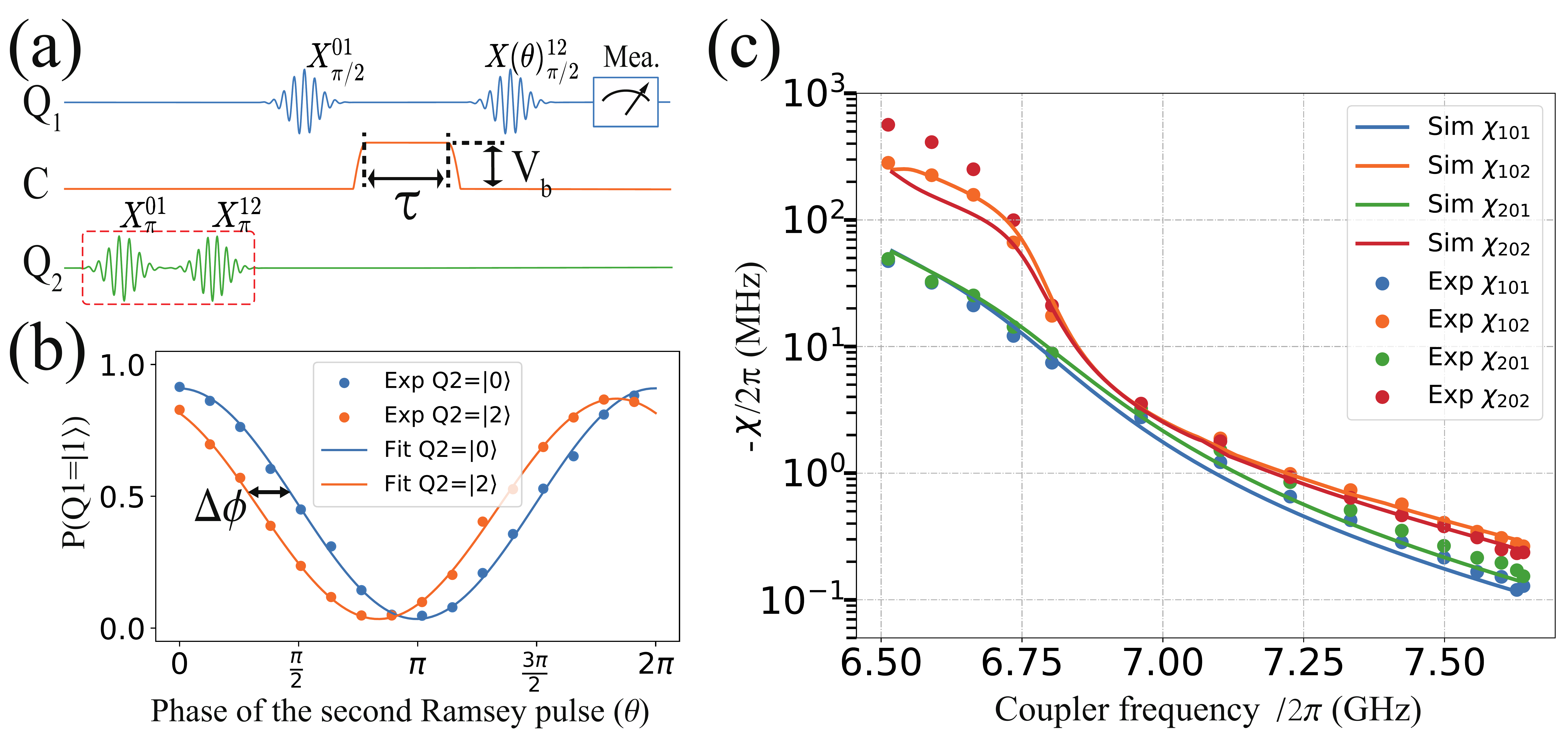}
    \caption{\label{fig:chi}Tunable cross-Kerr coupling between qutrits. (a) Pulse sequence of a Ramsey-like experiment on $Q_1$ conditioned by the state of $Q_2$ to extract the cross-Kerr coupling $\chi_{102}$. Two $\pi/2$ pulses are applied to $Q_1$. The first one generates a rotation around the $X$ axis in the {$|0\rangle$,$|1\rangle$} subspace and the second pulse corresponds to a rotation in the {$|1\rangle$,$|2\rangle$} subspace around an axis with an angle $\theta$ off the $X$ axis. Between the two pulses applied to $Q_1$, a square-shaped flux pulse with amplitude $V_b$ and duration $\tau$ is applied to the coupler, which induces a phase shift to be captured by $Q_1$. Depending on the magnitude of $V_b$, $\tau$ is set to different values ranging from 1 to 2000 ns (see supplemental material for details). Prior to the above three pulses, $Q_2$ is prepared into the second or the ground states (with or without applying the pulses enclosed by the red dashed lines). The conditional phase accumulated on $Q_1$ is measured as a function of $\theta$. (b) Experimental data (markers) are fitted by a sinusoidal function (solid lines) to extract the phase difference $\Delta \phi $ between the cases of $Q_2$ being prepared into the second excited (orange) and the ground (blue) states. The effective cross-Kerr coupling $\chi_{102}$ can be derived with the relation $\Delta\phi=\chi_{102} \tau$ at this fixed $V_b$ condition. Similar measurements are performed to obtain $\chi_{101},\chi_{201}$, and $\chi_{202}$. (c) Repeated experiments in (b) with different pulse amplitudes $V_b$, and combination the mapping relation pulse amplitudes $V_b$ to coupler frequency $\omega_c$ ( not shown here), we obtain four $\chi_{i0j}$(markers) depends on $\omega_c$ for two qutrits, which is in agreement with numerical results quite well(corresponding same color lines). We choose coupler sweet point $\omega_c\approx 7.640$ GHz as the idling point to keep small residual couplings in this work. }
    \end{figure}
    
    For the study of two-qutrit gates in this work, it is convenient to consider the effective form of $H$ in Eq.(\ref{eq1}) truncated to the computational subspace: $H_{eff}=\sum_{ij}\Tilde{\omega}_{i0j} |i0j\rangle \langle i0j|$ \cite{Yuan2020prl,ni2021scalable}, where $|i0j\rangle$ and $\Tilde{\omega}_{i0j}$ represent the eigenstates and eigenenergies of $H$, respectively. The tunable effective cross-Kerr interaction is defined as $\chi_{i0j}=\Tilde{\omega}_{i0j}-(\Tilde{\omega}_{i00}+\Tilde{\omega}_{00j})$ and is the consequence of interactions among the higher levels of the two qutrits. Both $\Tilde{\omega}_{i0j}$ and $\chi_{i0j}$
    are dependent on  $\omega_c$, the frequency of the tunable coupler. A nonzero $\chi_{i0j}$ results in an extra phase for the state of $|i0j\rangle$ relative to the phases accumulated on $|i00\rangle$ and $|00j\rangle$. Taking $|102\rangle$ as an example, its numerically simulated eigenenergy as a function of $\omega_c$ is plotted as the blue curve in Fig.~\ref{fig:devices}(d). As a comparison, the sum of $\Tilde{\omega}_{002}$ and $\Tilde{\omega}_{100}$ is plotted as the orange curve. When the coupler is far detuned from both qutrits, the blue and orange curves are nearly indistinguishable, indicating a negligible $\chi_{102}$ ($\sim 100$ kHz). However, $\chi_{102}$ increases to well above 100 MHz as the coupler is tuned closely to the qutrits due to the strong compelling interactions of multiple higher eigenstates. A high on-off ratio for $\chi_{i0j}$ can thus be achieved by varying the frequency of the coupler. 

    We first experimentally characterize the four cross-Kerr couplings $\chi_{i0j}$ $(i,j=1,2)$ using a conditional Ramsey-like experiment. The pulse sequence for measuring $\chi_{102}$, for example, is shown in Fig.~\ref{fig:chi}(a). The conditional phase difference $\Delta \phi=\chi_{102}\tau$ between $Q_2$ being in the states of $|0\rangle$ and $|2\rangle$ is the cumulative entangled phase of $|102\rangle$ during the Ramsey experiment \cite{Yuan2020prl}. To suppress the edge effect, we measure two conditional phases $\Delta \phi_1$ and $\Delta \phi_2$ corresponding to two different durations of $\tau_1$ and $\tau_2 $ with the same pulse amplitude. A more accurate $\chi_{102}$ can thus be obtained by $\chi_{102}=(\Delta \phi_1 - \Delta \phi_2)/(\tau_1-\tau_2)$. 

    By using the correspondence between the pulse amplitude and the coupler frequency, we plot in Fig.~\ref{fig:chi}(c) the four effective cross-Kerr couplings measured in the above conditional Ramsey experiment. The dynamic range of strength of these couplings spans more than two orders of magnitude, from a few hundred kHz to about 100 MHz, enabling fast two-qutrit gate operations as well as a small residual coupling. The results of $\chi_{101},\chi_{102}$ and $\chi_{201}$ are in good agreement with numerical simulations using our device parameters. The numerical simulation of $\chi_{202}$ quantitatively deviates from the experimental result at large values of $\chi_{202}$. Such a deviation is expected since the numerical simulation is based on the approximate Hamiltonian in Eq.~(\ref{eq1}), which assumes fixed values for the anharmonicity $\alpha_i$ and the coupling $g_{ij}$. More accurate simulation should take into account the frequency dependence of both quantities \cite{Didier2018pra}.
    
    \begin{figure}
    \includegraphics[width=0.47\textwidth]{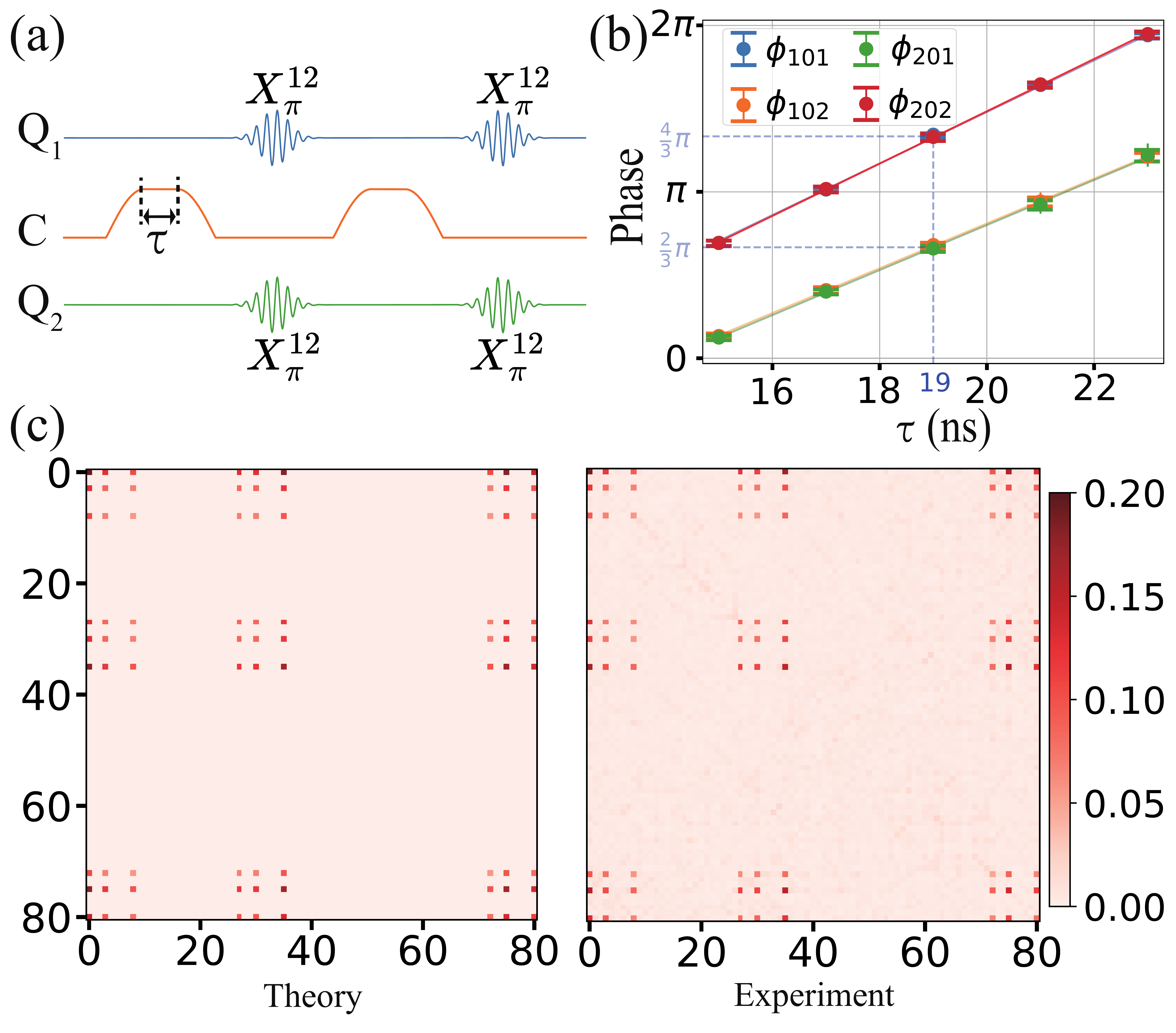}
    \caption{\label{fig3}Realization of two-qutrits Cphase gate. (a) The complete gate sequence consists of two identical pulses controlling the coupler frequency and four single-qutrit $\pi$ rotations in their {$|1\rangle$,$|2\rangle$} subspaces. The pulses applied to the coupler have flat tops with a duration of $\tau$ and two half-sine shaped edges. Each edge has a duration of 50 ns \cite{SupplementalMaterial}, time axis is not to scale. (b) Phases accumulated on different states as a function of $\tau$. Experimental results (markers) are fitted linearly with the error bars indicating one standard deviation. Notice that the two fitting lines are not parallel to each other and they reach $\frac{2}{3}\pi$ and $\frac{4}{3}\pi$ at $\tau$ = 19 ns, where the desired phase accumulation for the Cphase gate is fulfilled. (c) Full QPT results of the Cphase gate. Left and right panels are the ideal and experimental quantum process matrix $|\chi_{mn}|$. Coordinates on both axes number the orthogonal basis operators $E_{i}$ \cite{SupplementalMaterial}. }
    \end{figure}
    
    The large tunability of $\chi_{i0j}$ over the coupler frequency makes it possible to realize highly efficient conditional phase gates by engineering the pulse applied to the coupler. Here, we experimentally demonstrate a two-qutrit conditional phase gate as an extended version of the two-qubit conditional phase gate \cite{PhysRevXBlok}: $U_{Cphase}=e^{i\frac{2}{3}\pi}(|101\rangle \langle101 |+|202\rangle \langle 202 |) + e^{-i\frac{2}{3}\pi}(|102\rangle \langle102 |+|201\rangle \langle 201 |) + \sum_{i~or~j=0}|i0j\rangle \langle i0j |$. The previous schemes of two-qutrit $U_{Cphase}$ gate rely on a small and fixed cross-Kerr interaction and are implemented by interspersing the time evolution under the native cross-Kerr Hamiltonian with multiple single-qutrit gates \cite{PhysRevXBlok,Morvan2021PRL}. On the contrary, our scheme uses a tunable cross-Kerr interaction with a large on/off ratio so that the efficiency is much improved. The pulse sequence is shown in Fig.~\ref{fig3}(a). Two identical flux pulses are applied to the coupler while single-qutrit gates $X_{\pi}^{12}$ are implemented on both qutrits before and after the second flux pulse, which guarantees that $\phi_{101}=\phi_{202}$ and $\phi_{102}= \phi_{201}$. For a flux pulse with fixed amplitude and edge length, the four phases accumulated on $|101\rangle$, $|202\rangle$, $|201\rangle$, and $|102\rangle$ are controlled by setting the duration of the flat top of the flux pulse, as shown in Fig.~\ref{fig3}(b). When appropriate parameters are chosen so that $\phi_{101}(\phi_{202})$ mod $2\pi = 2\pi/3 $ and $\phi_{102}(\phi_{201})$ mod $2\pi  = 4\pi/3 $, a two-qutrit $U_{Cphase}$ gate is realized.

    Fidelity of the realized gate is characterized by a standard quantum process tomography (QPT) method \cite{Brien2004prl}. The full quantum process matrix $\chi_{exp}$ reconstructed experimentally is compared to the theoretically expected process matrix $\chi_{0}$ as shown in Fig.~\ref{fig3}(c). The gate fidelity is determined by $F = Tr(\chi_{exp}\chi_{0})= 89.3(\pm1.9)\%$, this result including errors of imperfect single qutrit gates for states preparation and tomography rotation. Numerical simulations using the open-source QuTiP package \cite{JOHANSSON20131234}, with or without decoherence, yield a gate fidelity of $91.2\%$ and $99.7\%$, respectively, here the single qutrit gate is simulated with optimal control technique derivative removal adiabatic gates (DRAG) \cite{Motzoi}. The effective decoherence time is calculated by averaging the decoherence time of qutrits along the coupler frequency trajectory during the total gate operation. For the specific two-qutrit gate trajectory in our work, the decoherence time are $T_1^{01}=27.1(16.8)\mu s $, $T_{2}^{01}=16.5(10.9)\mu s $, $T_1^{12}=13.9(8.9)\mu s $, $T_{2}^{12}=12.3(8.3)\mu s $ and $T_{2}^{02}=11.64(11.62)\mu s $ for $Q_{1}(Q_{2})$, detailed discussion is included in ref.\cite{SupplementalMaterial}. These results indicate that decoherence is the main source of error here, whereas other factors such as sates leakage and single qutrit gate infidelity are relatively small. 
    \begin{figure}
    \includegraphics[width=0.47\textwidth]{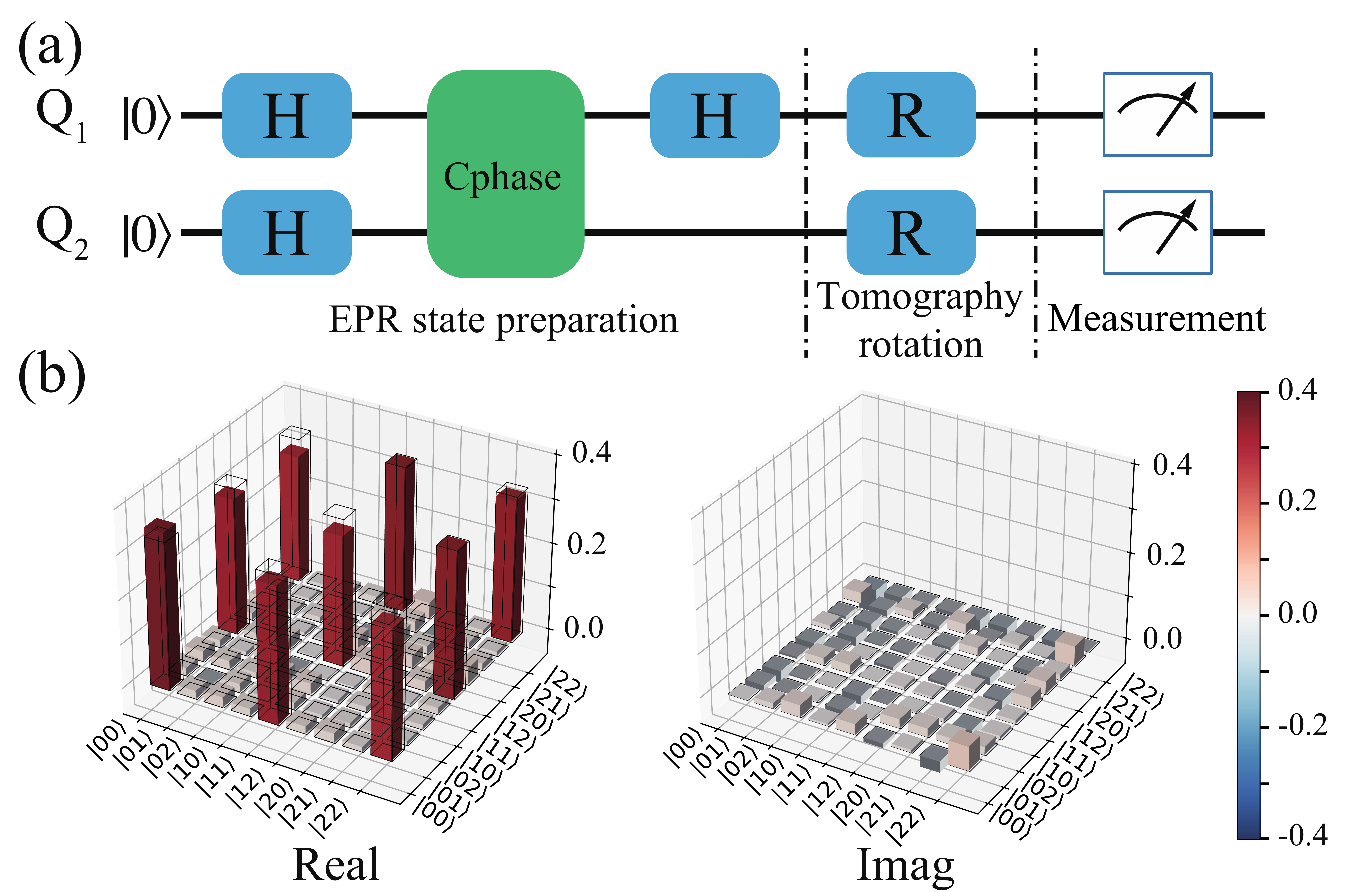}
    \caption{\label{fig4}Preparation of two-qutrit EPR state. (a) Quantum circuit for generating a two-qutrit EPR state using the Cpahse gate (green box) and single-qutrit Hadamard gates (blue boxes), followed by a QST characterization. (b) Real and imaginary parts of the density matrix of the two-qutrit EPR state. Solid bars and black outlines are experimental and ideal results, respectively.}
    \end{figure}
    
    To further benchmark the performance of the above two-qutrit Cphase gate, we use it to prepare a maximally entangled two-qutrit EPR state:
    \begin{equation}
    |\Psi_{EPR}\rangle = \frac{1}{\sqrt{3}} (|00\rangle + |11\rangle + |22\rangle)
    \end{equation}
    Similar to the case of two qubits, the two-qutrit EPR state can also be prepared by combining a two-qutrit Cphase gate with local single-qutrit Hadamard gates, as shown in Fig.~\ref{fig4}.(a). A standard quantum state tomography (QST) method is employed to reconstruct the state density matrix $\rho$. The results are shown in Fig.~\ref{fig4}.(b). The EPR state fidelity $F = \langle \Psi_{EPR}|\rho|\Psi_{EPR} \rangle$ is calculated to be $95.5(\pm2.0)\%$.
    
    In summary, we propose and experimentally demonstrate a new scheme of implementing two-qutrit Cphase gate. This scheme utilizes a tunable coupler to control the cross-Kerr coupling between the two involved qutrits so that a large on/off ratio (about a few hundred) can be achieved. Such a design can largely suppress errors due to the cross-talk among qutrits induced by a non-negligible residual coupling. In addition, the maximum accessible coupling makes fast two-qutrit gates possible. We demonstrate this scheme on a superconducting quantum circuit where two transmons of fixed frequencies serve as qutrits and another transmon with a tunable frequency controls their coupling. The resulted two-qutrit Cphase gate has a fidelity of $89.3(\pm1.9)\%$, mainly limited by decoherence. To show potential applications of this Cphase gate in quantum circuits, we use it to prepare a two-qutrit EPR state with a fidelity of $95.5(\pm2.0)\%$. We expect the gate performance can be further enhanced with pulse optimization \cite{chuji2021} and improved coherence \cite{Place2021,wang2022towards}, though careful pulse envelop need to take into consideration due to the more complicated energy spectra ( see supplemental material for more detail discussion \cite{SupplementalMaterial}). We anticipate that the two qutrits Cphase gate demonstrated at here will inspire further research in the field of ternary quantum information processing \cite{PhysRevAAlbarr,PhysRevXBlok,hrmo2022native,goss2022high}.

	\vspace{10pt}
	\noindent
	\textbf{Acknowledgements}\\
	{This work was supported by the Key-Area Research and Development Program of Guang-Dong Province (Grant No. 2018B030326001), the National Natural Science Foundation of China (Nos. U1801661 and 12004162), the Guangdong Provincial Key Laboratory (Grant No. 2019B121203002), the Science, Technology and Innovation Commission of Shenzhen Municipality (JCYJ20170412152620376, KYTDPT20181011104202253), and the Natural Science Foundation of Guangdong Province (Grant No. 2017B030308003).}
	
	\vspace{10pt}
	\noindent

    Note added —We notice a concurrent development of a similar work to implement Qutrit Entangling Gates with stark drives \cite{goss2022high}.

	\bibliography{TwoQutritsGateMain}
	\bibliographystyle{apsrev4-1}
\end{document}


\title{Supplemental Material of "Experimental Realization of Two Qutrits Gate with Tunable Coupling in Superconducting Circuits"}
	
	\author{Kai Luo}
	\affiliation{Department of Physics, Harbin Institute of Technology, Harbin 150001, China}
	\affiliation{Department of Physics, Southern University of Science and Technology, Shenzhen 518055, China}
	\affiliation{Shenzhen Institute for Quantum Science and Engineering, Southern University of Science and Technology, Shenzhen 518055, China}
	\author{Wenhui Huang}
	\author{Ziyu Tao}
	\author{Libo Zhang}
	\author{Yuxuan Zhou}
	\affiliation{Department of Physics, Southern University of Science and Technology, Shenzhen 518055, China}
	\affiliation{Shenzhen Institute for Quantum Science and Engineering, Southern University of Science and Technology, Shenzhen 518055, China}
	
	\author{Ji Chu}
	\affiliation{Shenzhen Institute for Quantum Science and Engineering, Southern University of Science and Technology, Shenzhen 518055, China}
	
	\author{Wuxin Liu}
	\affiliation{Huawei 2012 lab}
	\author{Biying Wang}
	\affiliation{Huawei 2012 lab}
	\author{Jiangyu Cui}
	\affiliation{Huawei 2012 lab}
	
	\author{Song Liu}
	\author{Fei Yan}
	\affiliation{Shenzhen Institute for Quantum Science and Engineering, Southern University of Science and Technology, Shenzhen 518055, China}
	\affiliation{Guangdong Provincial Key Laboratory of Quantum Science and Engineering, Southern University of Science and Technology, Shenzhen, 518055, China}
	\affiliation{Shenzhen Key Laboratory of Quantum Science and Engineering, Southern University of Science and Technology, Shenzhen 518055, China}
	\affiliation{International Quantum Academy, Shenzhen, Guangdong, China}
	
	\author{Man-Hong Yung}
    \email{yung@sustech.edu.cn }
    \affiliation{Department of Physics, Southern University of Science and Technology, Shenzhen 518055, China}
	\affiliation{Shenzhen Institute for Quantum Science and Engineering, Southern University of Science and Technology, Shenzhen 518055, China}
	\affiliation{Huawei 2012 lab}
	\affiliation{Guangdong Provincial Key Laboratory of Quantum Science and Engineering, Southern University of Science and Technology, Shenzhen, 518055, China}
	\affiliation{Shenzhen Key Laboratory of Quantum Science and Engineering, Southern University of Science and Technology, Shenzhen 518055, China}
	\affiliation{International Quantum Academy, Shenzhen, Guangdong, China}
	
	\author{Yuanzhen Chen}
	\email{chenyz@sustech.edu.cn}
	\affiliation{Department of Physics, Southern University of Science and Technology, Shenzhen 518055, China}
	\affiliation{Shenzhen Institute for Quantum Science and Engineering, Southern University of Science and Technology, Shenzhen 518055, China}
	\affiliation{Guangdong Provincial Key Laboratory of Quantum Science and Engineering, Southern University of Science and Technology, Shenzhen, 518055, China}
	\affiliation{Shenzhen Key Laboratory of Quantum Science and Engineering, Southern University of Science and Technology, Shenzhen 518055, China}
	\affiliation{International Quantum Academy, Shenzhen, Guangdong, China}
	
	\author{Tongxing Yan}
	\email{yantx@sustech.edu.cn}
	\affiliation{Shenzhen Institute for Quantum Science and Engineering, Southern University of Science and Technology, Shenzhen 518055, China}
	\affiliation{Guangdong Provincial Key Laboratory of Quantum Science and Engineering, Southern University of Science and Technology, Shenzhen, 518055, China}
	\affiliation{Shenzhen Key Laboratory of Quantum Science and Engineering, Southern University of Science and Technology, Shenzhen 518055, China}
	\affiliation{International Quantum Academy, Shenzhen, Guangdong, China}
	
	\author{Dapeng Yu}
	\affiliation{Department of Physics, Southern University of Science and Technology, Shenzhen 518055, China}
	\affiliation{Shenzhen Institute for Quantum Science and Engineering, Southern University of Science and Technology, Shenzhen 518055, China}
	\affiliation{Guangdong Provincial Key Laboratory of Quantum Science and Engineering, Southern University of Science and Technology, Shenzhen, 518055, China}
	\affiliation{Shenzhen Key Laboratory of Quantum Science and Engineering, Southern University of Science and Technology, Shenzhen 518055, China}
	\affiliation{International Quantum Academy, Shenzhen, Guangdong, China}
	
	\date{\today }
    
	\vskip 0.5cm
	\maketitle
	

	\section{Superconducting qubits processer}
	There are eight transmon type superconducting qubits on our devices but only two of them and a coupler are used in this work. Processer design is the same as \cite{chu2021scalable}. Qutrits frequency and other key parameters are shown in TABLE.S\ref{tab:device}.
    
    \begin{table}[ht]
    \caption{\label{tab:device} Qutrits parameters }
    \renewcommand\arraystretch{1.3}
    \begin{tabular}{p{8cm}<{\centering} p{2cm}<{\centering} p{2cm}<{\centering}}
    \hline
    Parameters (GHz)&Q1&Q2\\
    \hline
    Qutrit 01 transition frequency $\omega_{01}/2\pi$ & 6.074 & 6.725\\
    Qutrit 12 transition frequency $\omega_{12}/2\pi$ & 5.818 & 6.489 \\
    Qutrit anharmonicity $\alpha/2\pi$ & -0.256 & -0.236 \\
    Qutrit-Coupler coupling $g_{ic}/2\pi$ & 0.0985 & 0.1065 \\
    Qutrit-Qutrit direct coupling $g_{12}/2\pi$ & \multicolumn{2}{c}{$\sim 0.005$} \\
    \hline
    Coupler maximum 01 transition frequency $\omega_{01}/2\pi$ & \multicolumn{2}{c}{$7.64$} \\
    Coupler anharmonicity $\alpha/2\pi$ & \multicolumn{2}{c}{$-0.31$} \\
    \hline
    \end{tabular}
    \end{table}

    \section{Cphase gate pulse shape}
    The coupler between two qutrits has two Josephson junction SQUID structure, by which we can adjust the frequency of the coupler through external flux. The residual longitudinal coupling between two qutrits can be adjust by the coupler frequency. In order for the accumulation of phases to increase linearly with time, we use a square shape pulse with adiabatic edge to control the frequency of the coupler. For simplicity, the half-sine shape is chosen for the adiabatic edge in the two qutrits Cphase gate pulse in Fig. 3(a) in the main text. The rising edge and the falling edge have the same length of time $\tau_1$ and the square shape length is $\tau_2$. A part of the control signal can be written as:
    \begin{equation}
    \begin{split}
    V(t)=\begin{cases} V_{max}* sin(\frac{\pi}{\tau_1}t),   & t<\tau_1\\
    V_{max},  &  \tau_1 \leq t < \tau_1+ \tau_2 \\
    V_{max}* sin(\frac{\pi}{\tau_1}(t-(\tau_1+ \tau_2))+\frac{\pi}{2}), &  \tau_1+\tau_2 \leq t
    \end{cases}
    \end{split}
    \end{equation}
    In this work, $\tau_1=50~ns$, $\tau_2=19~ns$, and the single qutrit gate length is $40~ns$.
    
    \section{Two-qturit Cphase gate Error analysis}
    Generally speaking, gate fidelity is limited by three error mechanisms: decoherence, control error and state leakage. Imperfect controllable coupling to the system results in coherence errors, and coherence coupling between computational subspace and non-computational subspace can lead to state leakage. 
    
    In the absence of decoherence, the gate fidelity can be improved with carefully optimize the coherence coupling process. For the specific two-qutrit gate in our work, the transmon is multi-level system with weak anharmonicity, the non-computational subspace exists, which consists of the higher levels than second-excited state of the two fixed-frequency transmons and all excited levels of coupler. The gate error of our two-qutrit gate mainly comes from the imperfect single-qutrit gates and the leakage during modulating the coupler frequency. The two-qutrit gate fidelity characterized with QPT is 0.997 when using DRAG technique\cite{Motzoi} for four single qutrits in our simulation and is 0.992 without optimizing by DRAG. By comparison, the simulation result for two-qutrit gate fidelity is also 0.997 with perfect single qutrit assured, which suggests the single qutrit gate error is negligible here. 
    
\begin{figure}[ht]
    \includegraphics[width=0.7\textwidth]{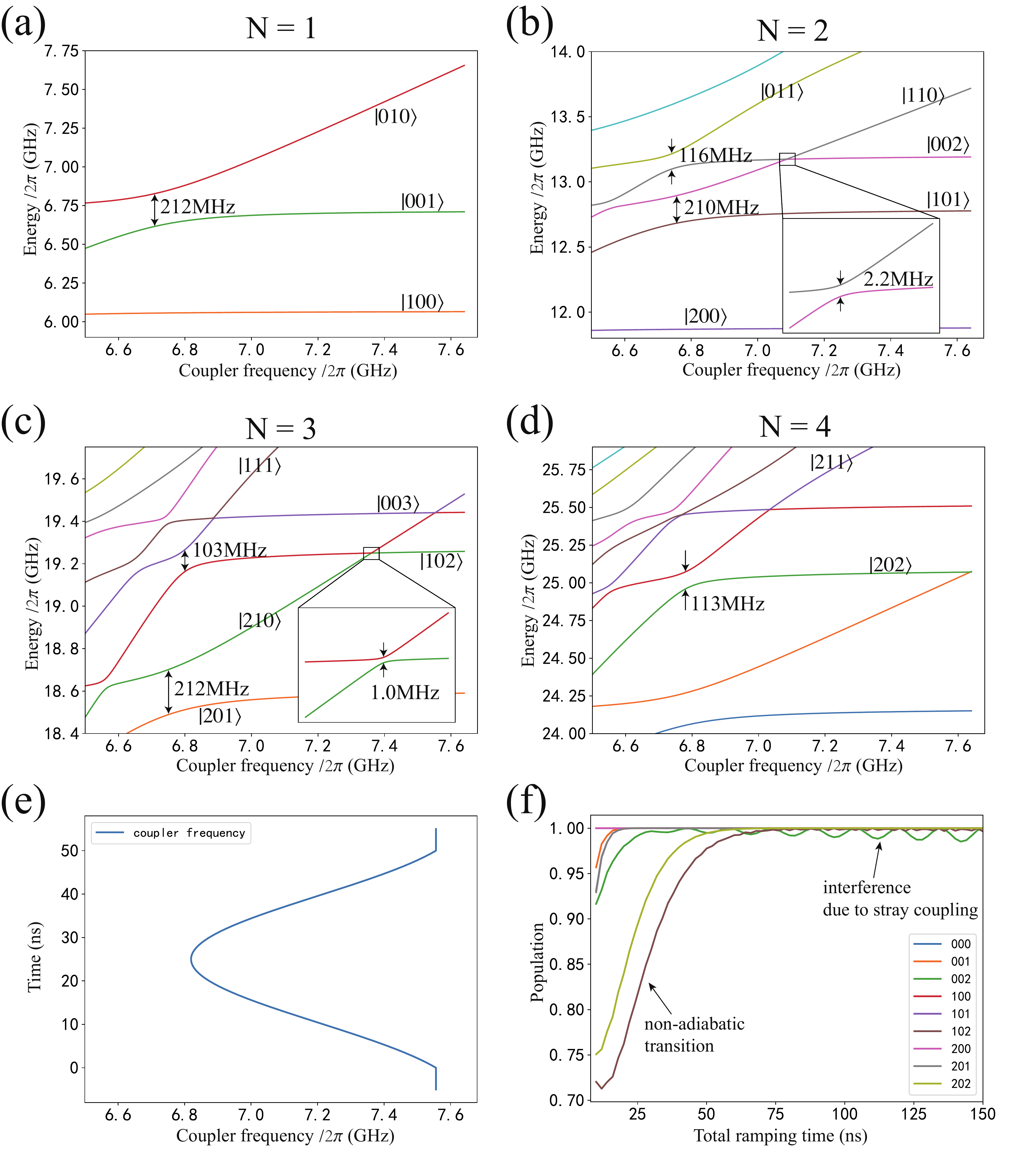}
    \caption{\label{SM_energyspectra}Energy-level spectra as a function of the coupler frequency, numerically calculated using Eq.(1) in main text. N in (a)-(d) denotes the total excitation number. All nine computational states for the two-qutrit gate are listed in the legend of (f). The gap between the computational and non-computational states is about $ 210~MHz$, as indicated for state $|001\rangle$ in (a) and $|101\rangle$ in (b). The gap is about $ 110~MHz$ for two-qutrit states $|002\rangle$, $|102\rangle$, and $|202\rangle$. Furthermore, a small gap about $1 \sim 2 ~MHz$ is observed for state $|002\rangle$ and $|102\rangle$,shown in the inserts of (b) and (c). (e) Coupler frequency as a function of thime during a ramping composed of two half-sine shaped flux pulses, with a total ramping time of $50~ns$. (f) Numerical simulation that characterizes leakage of each computational state. Initialize the system on nine different two-qutrit computational states and measure the residue probability after a period of ramp up and down trajectory illustrated in (e).}
    \end{figure}

    To see how the residual error of two-qutrit gate arises during tuning the coupler frequency, we need to dive into the detail of adiabatic process and energt-level spectra of the system Hamiltonian. The adiabatic condition between any two states, $|m\rangle$ and $|n\rangle$, can be written as:
    \begin{equation}
    \beta_{mn} = \arrowvert \frac{\hbar \langle m | \partial H/\partial t  |n \rangle }{(E_m-E_n)^2} \arrowvert  \ll 1, ~~~ m \ne n
    \end{equation}
    which implies the unwanted non-adiabatic transition depends on the slope of Hamiltonian (the slope of the coupler frequency trajectory in our case) and the energy gap between them. The energy spectra of EQ.1 in our manuscript as a function of the coupler frequency is shown in Fig.S\ref{SM_energyspectra}. The minimum gap for two-qubit computational states and non-computational states is $\sim 210~ MHz$, as indicated for states $|001\rangle$ and $|101\rangle$, while the gap is $\sim 110~ MHz$ for two-qutrit states $|002\rangle$, $|102\rangle$, and $|202\rangle$. The smaller gaps suggest that, for same circuit design, the ramp up for two-qutrit gate should be slower than the two-qubit gate to satisfy adiabatic condition well. 
    
    To show how the ramp up time affects the nonadiabatic leakage, we simulate all the nine computational states leakage after modulating one period of the coupler frequency by a half-sine flux pulse (remind that our two-qutrit gate process consists of tuning the coupler two times plus four single qutrit gate) by varying the total ramp up time. An example of the coupler-frequency vs time of $50~ns$ length (equal ramp up and down with $25~ns$) is plotted in Fig.S\ref{SM_energyspectra}(e), and the probability returning back to the initial nine computational states is shown in Fig.S\ref{SM_energyspectra}(f). The leakage errors for states $|002\rangle$, $|102\rangle$, and $|202\rangle$ are serious when the total ramp time (sum of up and down) is less than $80 ~ns$. In our experiment, we used a longer ramp time $100~ns$ for safety consideration.
  
\begin{figure}[ht]
    \includegraphics[width=0.5\textwidth]{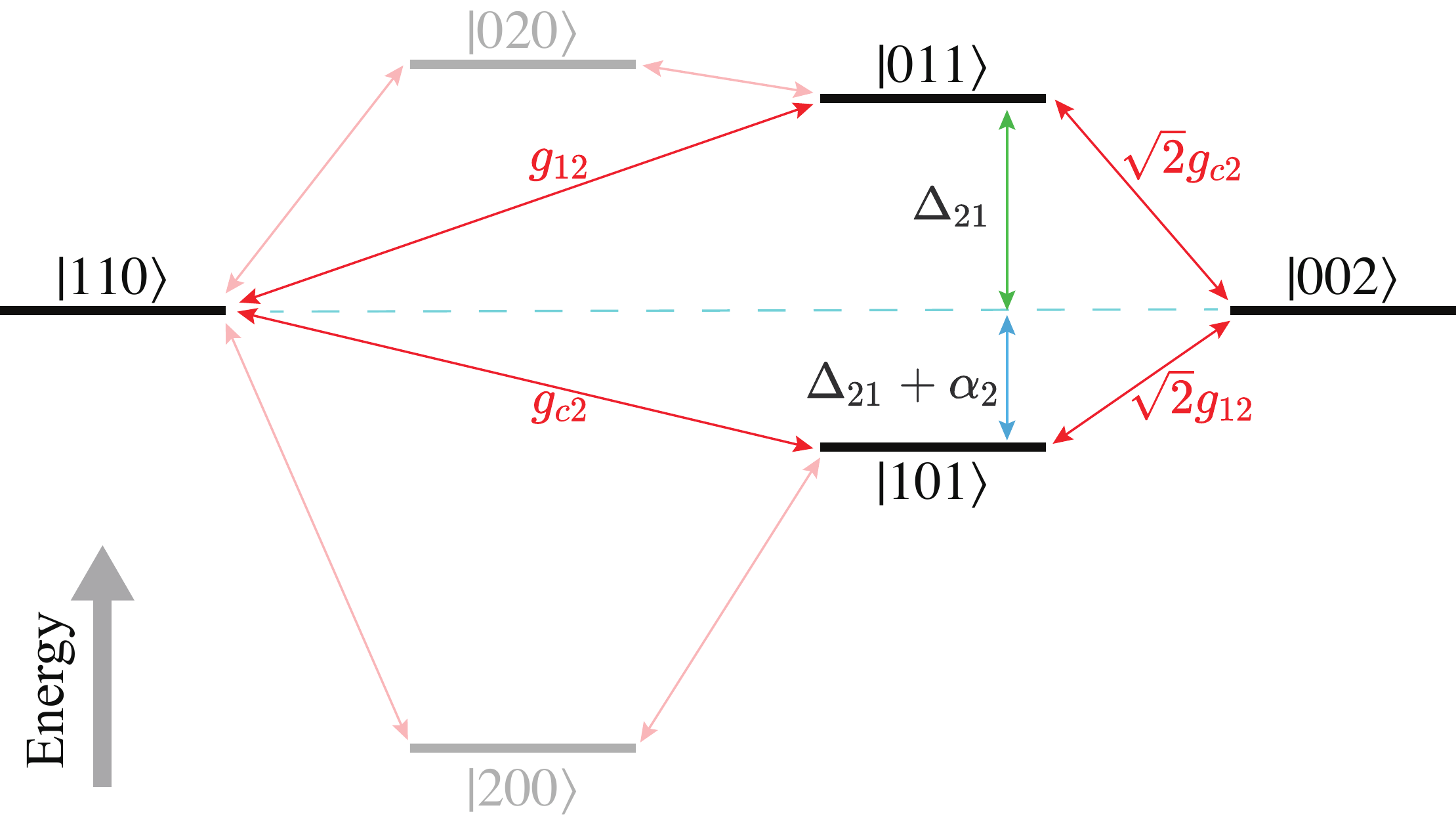}
    \caption{\label{SM_coupling110_002}Energy configurations for the effective three body interaction between $|002\rangle$ and $|110\rangle$. The lines denote the levels with distances in a vertical direction proportional to their energy differences. The double arrows denote the allowed transitions, with the bright red ones denoting the major contributing path.} 
    \end{figure}

    Furthermore, an unexpected small gap about $1 \sim 2 ~ MHz$ is observed for state $|002\rangle$  and $|102\rangle$  as zoom up in Fig.S\ref{SM_energyspectra}(c) and (d). These stray coupling rise from the effective three body interaction, which is a Raman process. We take state $|002\rangle$ as example to explain it. The small gap is observed when the state $|110\rangle$ is tuned to resonant with $|002\rangle$, noticing that this is a three body interaction and it's not explicit shown in our system Hamiltonian. To reveal how this happen, we plot the energy configurations for related levels as Fig.S\ref{SM_coupling110_002}. As a semi-quantitative estimation, we only consider the virtual exchange process through states $|011\rangle$ and $|101\rangle$, and ignore $|200\rangle$ and $|020\rangle$. Using the Schrieffer-Wolff transformation, one can derive the effective stray exchange coupling strength between state $|002\rangle$ and $|110\rangle$:

    \begin{equation}
    g_{002,110} \approx \frac{g_{110,011}g_{011,002}}{E_{110}-E_{011}} + \frac{g_{110,101}g_{101,002}}{E_{002}-E_{101}} 
    \approx \frac{g_{12}\sqrt{2}g_{c2}}{\Delta_{21}} + \frac{g_{c2}\sqrt{2}g_{12}}{-(\Delta_{21}+\alpha_2)} \approx 0.6 ~ MHz
    \end{equation}

    \noindent where $g_{ij}$ is the coupling strength between state $i$ and $j$, $\Delta_{21} \equiv \omega_{q2}^{01}-\omega_{q1}^{01}$ is the detuning of two qutrits and $\alpha_2$ is the anharmonicity of Q2. This simple approximation is consist with numerical calculation $2.2/2=1.1 ~MHz$.
    To keep the system at the wanted computational state and suppress leakage, sweep through this crossing should be fast so that is nearly a fully diabatic process. Nonnegligible interference oscillation for state $|002\rangle$ is observed in Fig.S\ref{SM_energyspectra}(f) due to its stray coupling is stronger and the slope of coupler frequency is relatively smaller near this frequency region. A possible way to further reduce the error is to optimize the waveform so that the oscillation due to stray coupling is destructive interference and the computational state remains in itself when the whole gate process is done.

    \section{ Decoherence times}
\begin{figure}[ht]
    \vspace{-1em}  
    \includegraphics[width=0.45\textwidth]{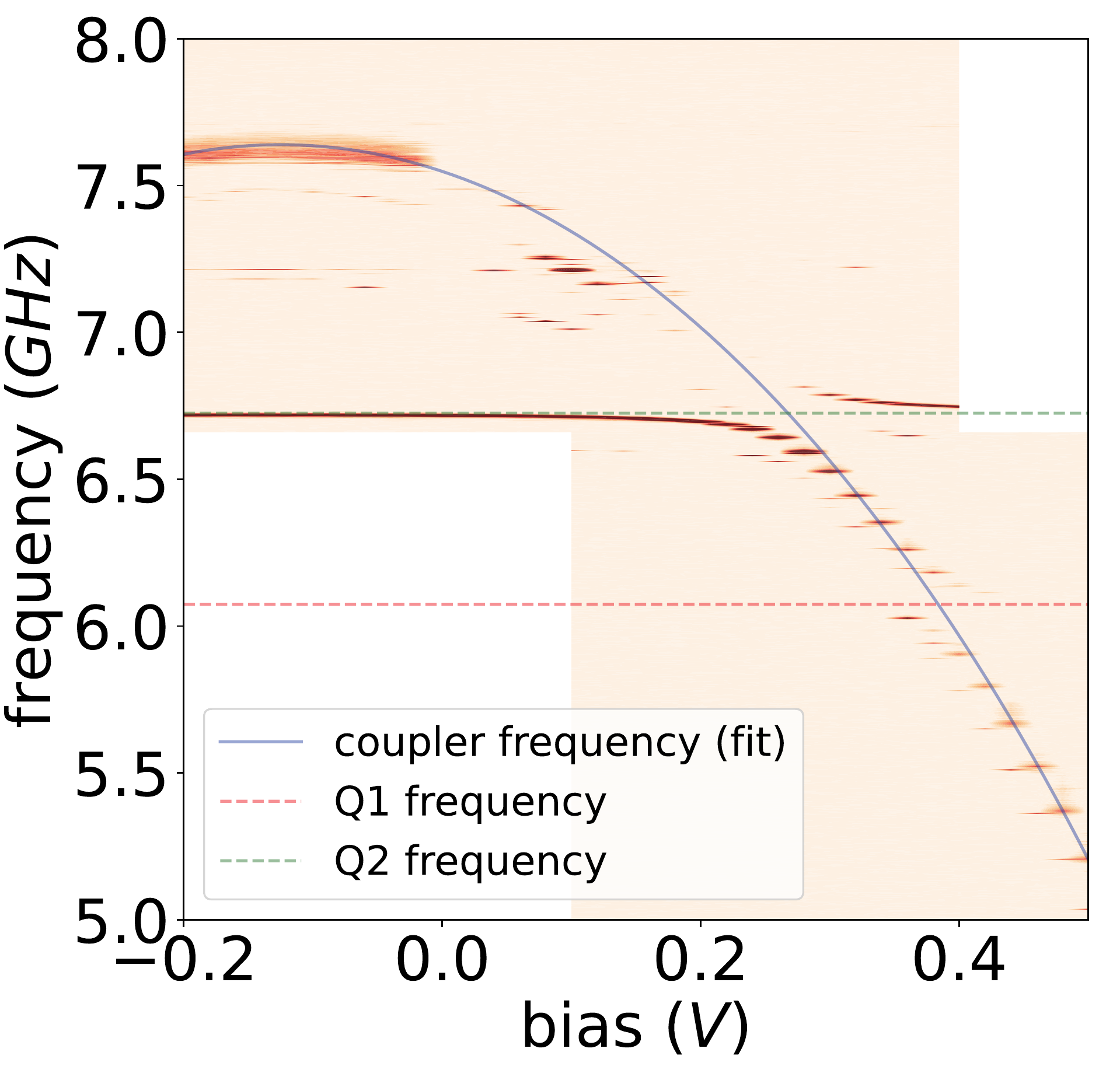}
    \caption{\label{SM_couplerfreqVSbias}Coupler frequency vs. bias amplitude. A long and strong drive is applied to the qutrit to measure the energy spectrum of the system at the same time as the Z pulse modulating coupler frequency is applied. The blue full line is the coupler frequency fitted as $7.95*\sqrt{cos(1.71*(bias+0.125))}-0.31$ and the red(green) dotted line is the frequency of $Q1^{01}(Q2^{01})$. Because of the strong interaction between the qutrits and the coupler, the energy spectrum splits when the coupler and the qutrits are close to each other. The minimum value of energy spectrum splitting is equal to two times of coupling strength (coupler-qutrit). }
    
    \vspace{-1em}  
    \end{figure}

\begin{figure}[ht]
    \vspace{-1em} 
    \includegraphics[width=0.6\textwidth]{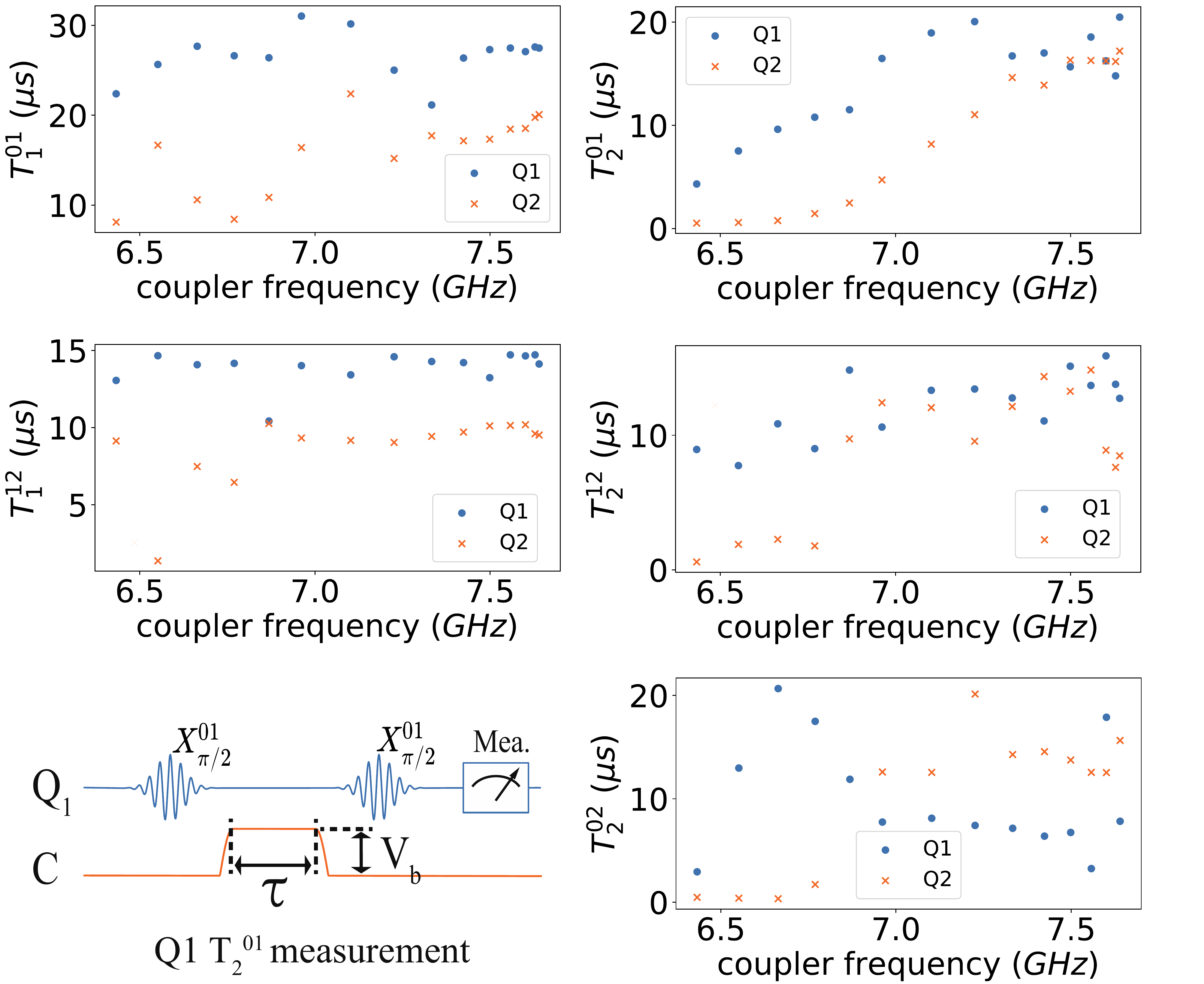}
    \caption{\label{SM_decoherencetime}Qutrits decoherence time vs. coupler frequency. The coupler is biased to different frequency while measuring qutrits' decoherence time. For example, a square pulse with adiabatic edge is inserted into the Ramsey-like experiment for measuring Q1 $T_{2}^{01}$. }
    \vspace{-1em}
    \end{figure}

    The coupler between two qubit has no readout resonator. In order to estimate the effect of decoherence on the experiment, we measure the relationship between the coupler frequency and flux pulse amplitude (bias) and the effective decoherence times of both qutrits when the coupler at different frequencies, which are shown in Fig.S\ref{SM_couplerfreqVSbias} and Fig.S\ref{SM_decoherencetime}, respectively. $T_2$ effectively describes both decoherence processes including energy dissipation time $T_1$ and pure dephasing time $T_{\phi}$. According to the trajectory of the coupler, the effective decoherence times is calculated by averaging the decoherence time at each moment: $T_{eff}=(\int_{0}^{\tau}T_{decoherence}(t) dt)/\tau$, $\tau$ is the total duration of the Cphase gate. As a result, the effective decoherence times of Q1(Q2) are $T_1^{01}=27.1(16.8)~\mu s $, $T_{2}^{01}=16.5(10.9)~\mu s $, $T_1^{12}=13.9(8.9)~\mu s $, $T_{2}^{12}=12.3(8.3)~\mu s $ and $T_{2}^{02}=11.6(11.6)~\mu s $.
    

    \section{ Quantum process tomography}
\begin{figure}[ht]
    \includegraphics[width=0.6\textwidth]{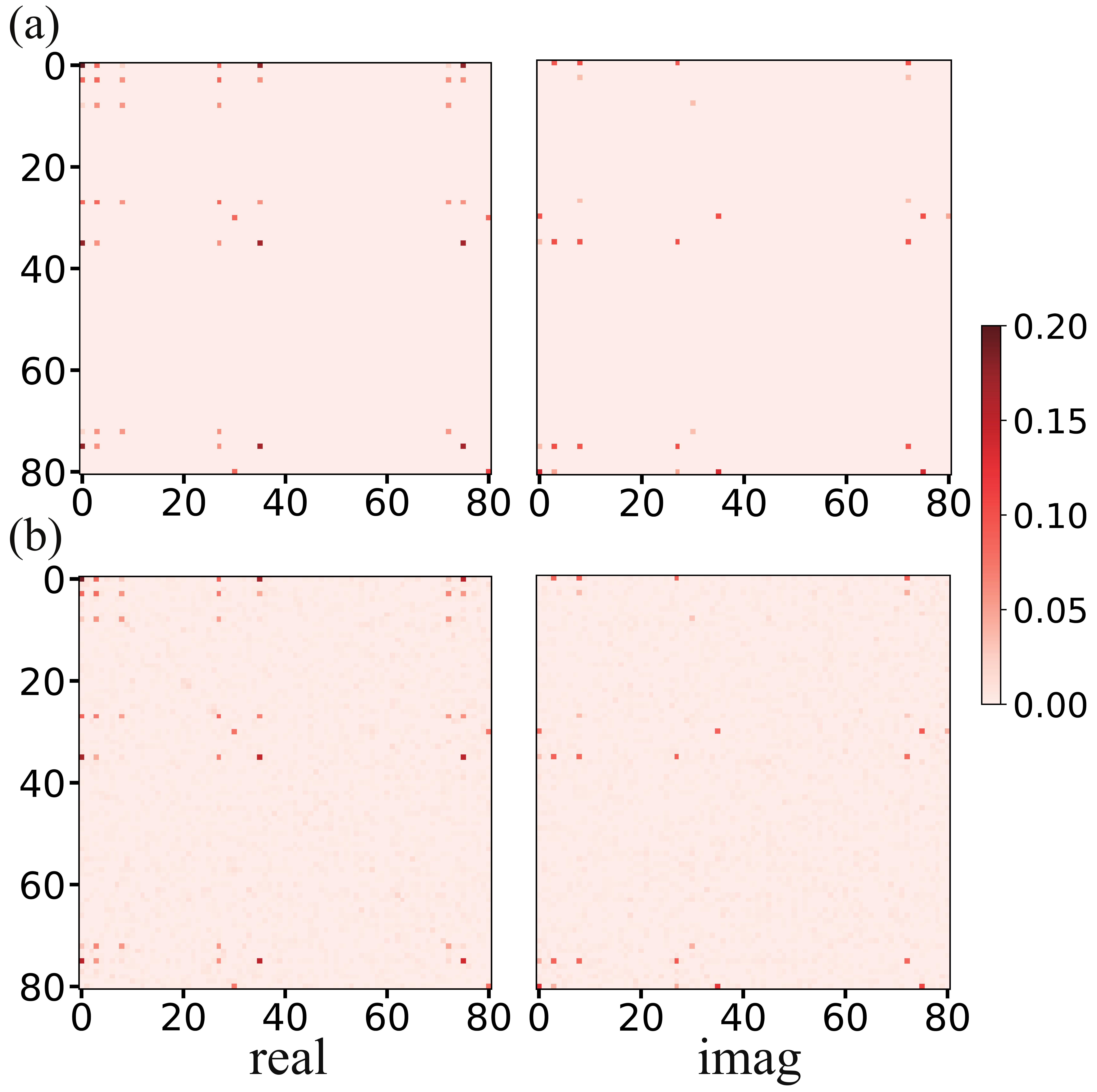}
    \caption{\label{SM_QPTresult}The full QPT result of two-qutrit Cphase gate. (a) is the ideal result and (b) is the measured result. The experimental gate fidelity is 89.3\%, note that this result includes the state preparation and measurement (SPAM) errors }
    \end{figure}

\begin{figure}[ht]
    \includegraphics[width=0.6\textwidth]{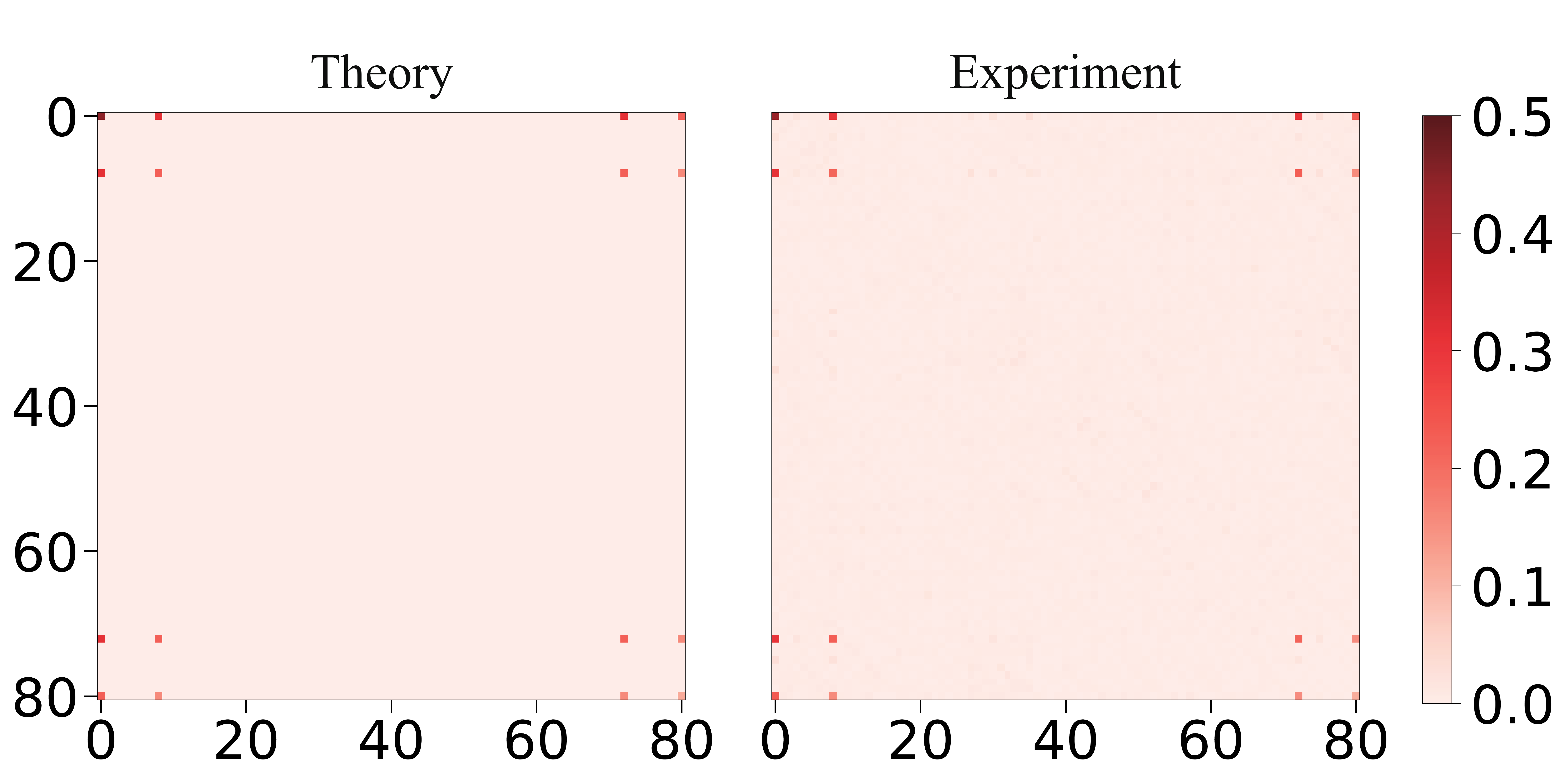}
    \caption{\label{SM_I_qpt}The QPT result of two-qutrit identity gate. Only the absolute value of $\chi$ is shown, because the imaginary part of both is almost zero.The experimental two-qutrit identity gate QPT fidelity is 98.5\%, the 1.5\% infidelity indicates the state preparation and measurement (SPAM) errors is pretty small compared to the two-qutrit Cphase gate of our work (about 10\%).}
    \end{figure}

    Quantum process tomography is used to characterizing the two qutrits Cphase gate. Firstly, initialize qutrits to the following 81 states respectively:$ \{ |0\rangle,~|1\rangle,~|2\rangle,~(|0\rangle+|1\rangle)/\sqrt{2},~(|1\rangle+|2\rangle)/\sqrt{2},~(|0\rangle+|2\rangle)/\sqrt{2},~(|0\rangle-i|1\rangle)/\sqrt{2},~(|1\rangle-i|2\rangle)/\sqrt{2},~(|0\rangle-i|2\rangle)/\sqrt{2}  \}\bigotimes^2$; then apply the Cphase gate on two qutrits; after that, for each initial state, the following 81 rotation gates are used for reconstructing the density matrix (also called quantum state tomography):$\{ I,~X_{\frac{\pi}{2}}^{01},~Y_{\frac{\pi}{2}}^{01}$,
    $~X_{\pi}^{01},~X_{\frac{\pi}{2}}^{12},~Y_{\frac{\pi}{2}}^{12},~(Y)_{\pi}^{12} X_{\frac{\pi}{2}}^{01} (-Y)_{\pi}^{12},~(Y)_{\pi}^{12} Y_{\frac{\pi}{2}}^{01} (-Y)_{\pi}^{12},~(Y)_{\pi}^{12} X_{\pi}^{01} (-Y)_{\pi}^{12} \}\bigotimes^2$, where the subscript indicate the rotation angle and the superscript indicate the operating subspace, (-Y) means the rotation axis is the minus Y-axis. Finally, the “Cphase gate” map can be written as: $\rho_f = \sum_{m,n} \chi_{mn} E_m \rho_i E_n^\dagger $, where the orthogonal basis operators for two-qutrits are $\{E_j\}\bigotimes^2$. The matrix representation of that for single-qutrit $\{E_j\}$ are:
    
    \begin{eqnarray}
    E_0=\sqrt{\frac{3}{2}}\begin{pmatrix}
    1 & 0 & 0\\
    0 & 0 & 0\\
    0 & 0 & 1\\
    \end{pmatrix},~~~
    &E_1=\sqrt{\frac{3}{2}}\begin{pmatrix}
        0 & 0 & 1\\
        0 & 0 & 0\\
        1 & 0 & 0\\
    \end{pmatrix},~~~
    &E_2=\sqrt{\frac{3}{2}}\begin{pmatrix}
        0 & 0 & -1\\
        0 & 0 & 0\\
        1 & 0 & 0\\
    \end{pmatrix}, \notag\\
    E_3=\sqrt{\frac{3}{2}}\begin{pmatrix}
    1 & 0 & 0\\
    0 & 0 & 0\\
    0 & 0 & -1\\
    \end{pmatrix},~~~
    &E_4=\sqrt{\frac{3}{2}}\begin{pmatrix}
    0 & 1 & 0\\
    1 & 0 & 0\\
    0 & 0 & 0\\
    \end{pmatrix},~~~
    &E_5=\sqrt{\frac{3}{2}}\begin{pmatrix}
    0 & -1 & 0\\
    1 & 0 & 0\\
    0 & 0 & 0\\
    \end{pmatrix}, \notag\\
    E_6=\sqrt{\frac{3}{2}}\begin{pmatrix}
    0 & 0 & 0\\
    0 & 0 & 1\\
    0 & 1 & 0\\
    \end{pmatrix},~~~
    &E_7=\sqrt{\frac{3}{2}}\begin{pmatrix}
    0 & 0 & 0\\
    0 & 0 & -1\\
    0 & 1 & 0\\
    \end{pmatrix},~~~
    &E_8=\sqrt{3}\begin{pmatrix}
    0 & 0 & 0\\
    0 & 1 & 0\\
    0 & 0 & 0\\
    \end{pmatrix} \notag
    \end{eqnarray}
    Both real and imaginary part of the ideal(a) and experimental(b) QPT result are shown in Fig.S\ref{SM_QPTresult}, which are used to get Fig.3 in the main text. For comparison, the ideal and experimental QPT result of I gate (do nothing) is shown in Fig.S\ref{SM_I_qpt}. The fidelity of experimental result is 98.5\%, which indicates the state preparation and measurement (SPAM) errors is about 1.5\%.
    
    \section{ The Lindblad Master equation}
    A commonly used method for calculating the dynamical evolution of a system with decoherence is the Lindblad master equation, which can be written as:
    \begin{eqnarray}
    \dot{\rho}(t)=-\frac{i}{\hbar}[H(t),\rho(t)]+\sum_n \frac{1}{2}[2C_n \rho(t) C^\dagger_n - \rho(t)C^\dagger C_n - C^\dagger_n C_n \rho(t)]
    \end{eqnarray}
    where $\rho$ is the system density matrix, $H$ is the system Hamiltonian with drive and $C_n$ are collapse operators. In this work, $C_n \in \{ \frac{1}{\sqrt{T_{1,Q1}^{01}}} a_1,~ \frac{1}{\sqrt{T_{1,Q2}^{01}}} a_2,~  \frac{2}{\sqrt{T_{\phi,Q1}^{01}}} a_1^\dagger a_1,~  \frac{2}{\sqrt{T_{\phi,Q2}^{01}}} a_2^\dagger a_2 \}$ where $a_i(a_i^\dagger)$ is the corresponding creation(annihilation) operator of the qutrit $i$ and $T_{\phi}^{01}$ is calculated by $1/T_{2}^{01}=1/(2T_{1}^{01}) + 1/T_{\phi}^{01}$. For simplicity, we use the effective decoherence time $T_{eff}$ here and omit the collapse operators of the coupler.
    
\begin{figure}[ht]
    \includegraphics[width=0.7\textwidth]{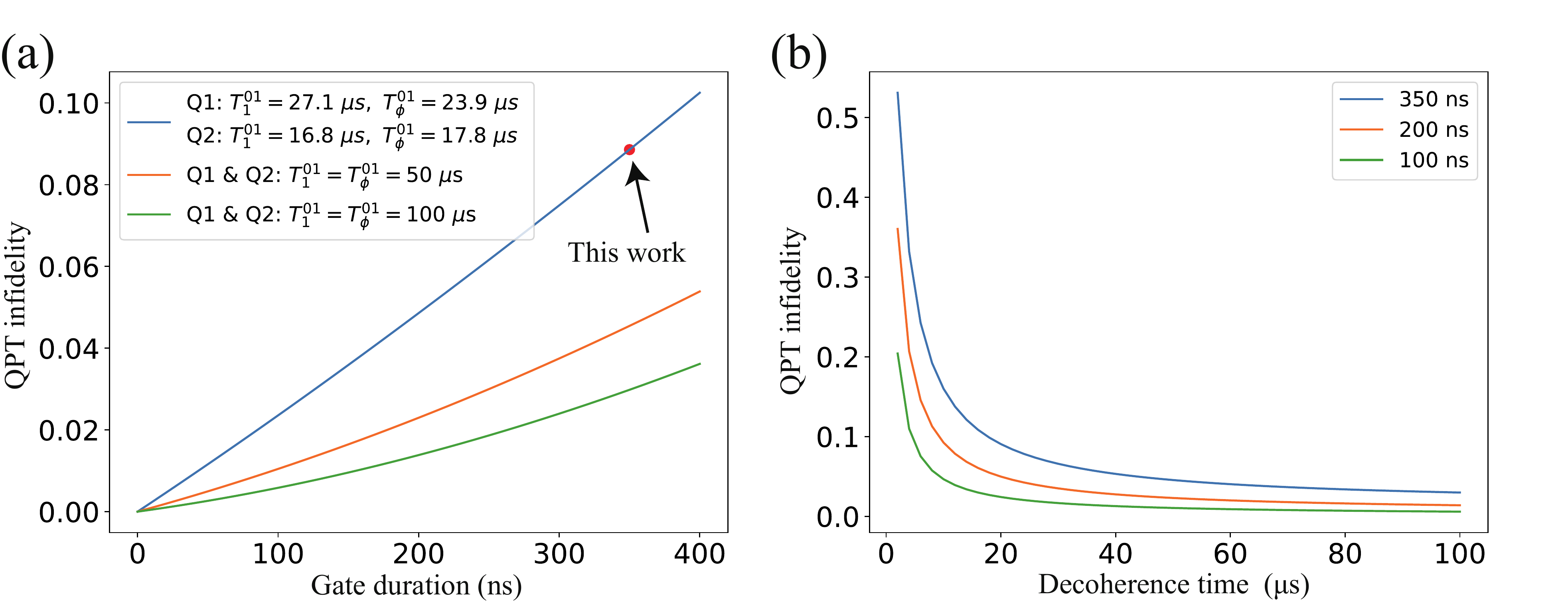}
    \caption{\label{SM_I_fidelity_Qlifetime}The numerical simulation result of I gate QPT infidelity with perfect state preparation and tomography. (a) Infidelity under fixed qutrits' lifetime and varied I gate duration (decoherence duration), where 50(100)~$\mu s$ denote $T_{1}^{01}=T_{\phi}^{01}=50(100)~\mu s$. The parameters of our device are in section.IV. The red dot indicate a gate duration of $350~ns$, the same as the Cphase gate. (b) Infidelity under fixed I gate duration and varied qutrits' lifetime. In our experiment, there $10~ns$ time buffer between flux pulse and $40~ns$ single qutrit gate, so total duration for Cphase gate is about $350~ns$ in this paper.}
    \end{figure}

    To estimate the upper limit of fidelity, we use the master equation to simulate the infidelity without SPAM of I gate as a function of gate duration for different decoherence times, and that under fixed gate duration and different decoherence time, which are shown as Fig.S\ref{SM_I_fidelity_Qlifetime}.
    
    \section{ Two qutrits readout}
\begin{figure}[ht]
    \includegraphics[width=0.5\textwidth]{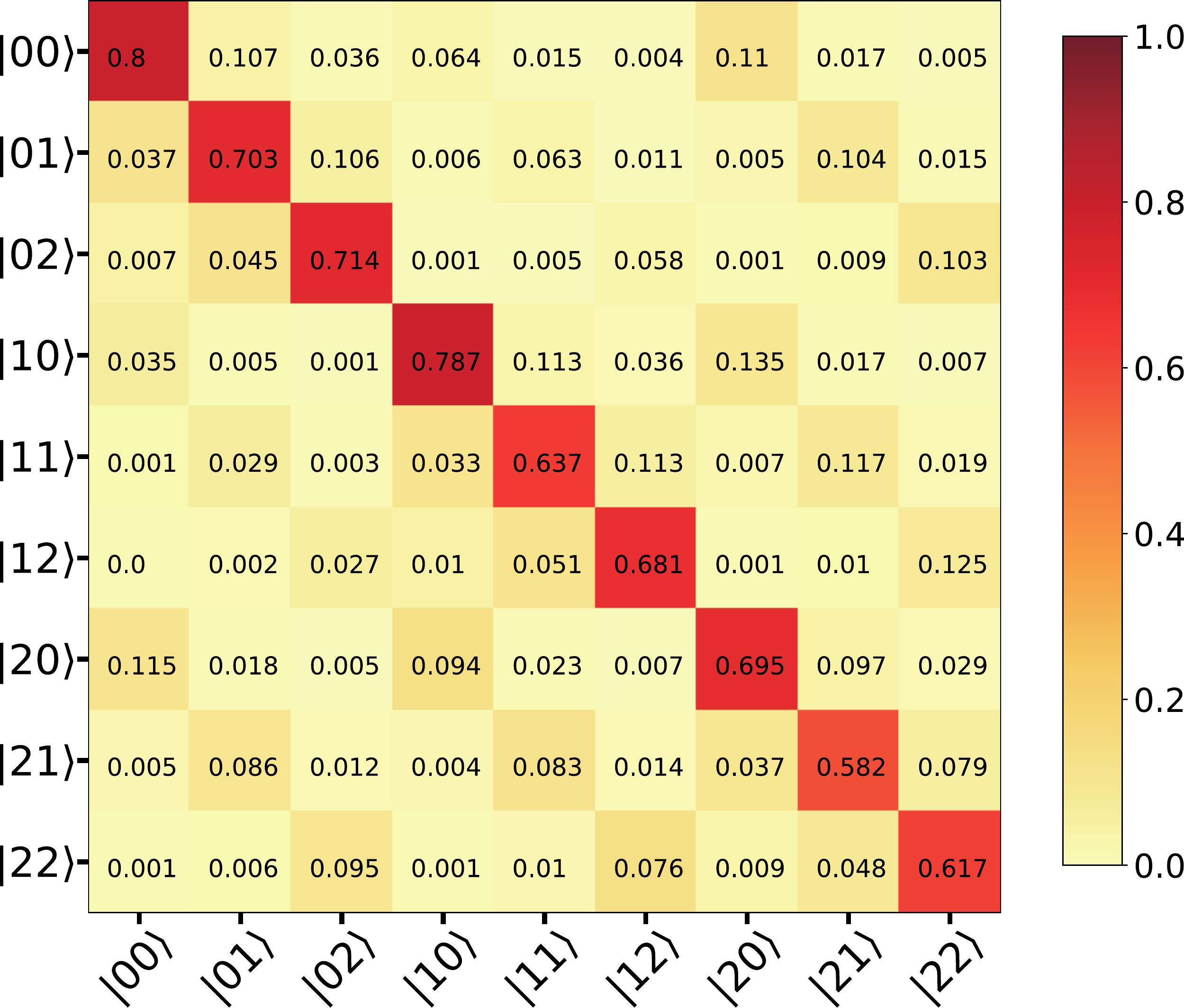}
    \caption{\label{SM_measurefidelity}The measure fidelity of nine two qutrits states. Single qutrit operations and readout are carried out when the coupler is far detunned from both qutrits (at the sweet point). We measure the assignment probability matrix $\mathcal{M}$ for mitigating the preparation and readout error.}
    \end{figure}

    The multiplexed dispersive readout scheme is used to distinguish two qutrits states. We prepare nine initial states and then measure the population of each state to get the assignment probability matrix $\mathcal{M}$, which is shown in Fig.S\ref{SM_measurefidelity}. In QPT experiment, the measured probability $P_m$ is multiplied by the inverse of the matrix to mitigate the measurement error, $P_{cor}=\mathcal{M}^{-1} \cdot P_m$.

	\bibliography{SupplementalMaterial}
	\bibliographystyle{apsrev4-1}